\documentclass[a4paper, 12pt]{article}

\usepackage{natbib}
\bibpunct{(}{)}{;}{a}{}{,}

\usepackage{graphicx}
\usepackage{amsfonts}
\usepackage{latexsym}
\usepackage{amsmath, amsthm, amssymb}
\usepackage{setspace}
\usepackage{url}
\usepackage{multirow}
\usepackage[margin=1in]{geometry}
\usepackage{epsfig}
\usepackage{psfrag}
\usepackage{rotating}

\makeatletter
\def\url@leostyle{%
  \@ifundefined{selectfont}{\def\UrlFont{\sf}}{\def\UrlFont{\small\ttfamily}}}
\makeatother
\theoremstyle{definition}
\newtheorem{thm}{Theorem}
\theoremstyle{definition}

\theoremstyle{definition}

\theoremstyle{definition}
\newtheorem{prop}{Proposition}
\theoremstyle{definition}

\theoremstyle{remark}

\theoremstyle{definition}
\newtheorem{defn}{Definition}

\newcommand{\E}{\mathbb{E}}
\newcommand{\R}{\mathbb{R}}
\newcommand{\var}{\mbox{var}}
\newcommand{\cov}{\mbox{cov}}
\newcommand{\rk}{\mbox{rk}}
\newcommand{\inner}[2]{\langle #1, #2 \rangle}

\newcommand{\gam}{\gamma}
\newcommand{\lam}{\lambda}

\newcommand{\vep}{\varepsilon}
\newcommand{\bvep}{\boldsymbol{\vep}}
\newcommand{\bbeta}{\boldsymbol{\beta}}

\newcommand{\bLam}{\boldsymbol{\Lambda}}
\newcommand{\bSig}{\boldsymbol{\Sigma}}

\newcommand{\bv}{\mathbf{v}}
\newcommand{\bx}{\mathbf{x}}
\newcommand{\by}{\mathbf{y}}
\newcommand{\bB}{\mathbf{B}}
\newcommand{\bI}{\mathbf{I}}

\newcommand{\bU}{\mathbf{U}}
\newcommand{\bV}{\mathbf{V}}

\newcommand{\cI}{\mathcal{I}}
\newcommand{\cL}{\mathcal{L}}

\def\wh{\widehat}
\def\wt{\widetilde}

\title{
Modelling and forecasting daily electricity load curves: a hybrid approach
\footnote{Partially supported by the EPSRC research grant EP/G026874/1.}
}

\author{
Haeran Cho\thanks{Department of Statistics, London School of Economics, UK.}
\and
Yannig Goude\thanks{\'{E}lectricit\'{e} de France, France.} 
\and 
Xavier Brossat\footnotemark[3]
\and 
Qiwei Yao{\footnotemark[2] $^{,}$\thanks{Guanghua School of Management, Peking University, China}}
}

\begin{document}

\date{}
\maketitle

\begin{abstract}
We propose a hybrid approach for the modelling and the short-term
forecasting of electricity loads.
Two building blocks of our approach are
(i) modelling the overall trend and seasonality by fitting a generalised additive model 
to the \emph{weekly} averages of the load, and 
(ii) modelling the dependence structure across consecutive \emph{daily} loads via curve linear regression.
For the latter, a new methodology is proposed for
linear regression with both curve response and curve regressors.
The key idea behind the proposed methodology is the dimension reduction based on 
a singular value decomposition in a Hilbert space, 
which reduces the curve regression problem to several ordinary (i.e. scalar) linear regression problems.
We illustrate the hybrid method using the French electricity loads between 1996 and 2009, 
on which we also compare our method with other available models including the EDF operational model.
\end{abstract}

\bigskip


\bigskip

\section{Introduction}
\label{sec:intro}

As electricity can be stored or discharged only at extra costs, 
it is an important task for electricity providers to model and forecast 
electricity loads accurately over short-term (from one day to one month ahead) 
or middle-term (from one month to five years ahead) horizons. 
The electricity load forecast is an essential entry of the optimisation tools
adopted by energy companies for power system scheduling. 
A small improvement in the load forecasting can bring in substantial benefits
in reducing the production costs as well as increasing the trading advantages, 
especially during the peak periods.  

The French energy company \'{E}lectricit\'{e} de France (EDF) manages a
large panel of production
units in France and in Europe, which include water dams, nuclear plants, wind turbines, coal and gas plants. 
Over the years, EDF has developed a very accurate load forecasting model 
which consists of complex regression methods coupled with classical time series techniques 
such as the seasonal ARIMA (SARIMA) model.
The model integrates a great deal of physical knowledge on the French electricity consumption patterns
that has been accumulated over 20 years,
such as the fact that the temperature felt indoors is more relevant than
the real temperature in modelling the electricity load. 
Furthermore, it includes exogenous information ranging from economic growth forecasts 
to different tariff options provided by the company.
The forecasting model in operation performs very well at present, 
attaining about 1\% mean absolute percentage error in forecasting over one day horizon.
However, it has a drawback in terms of its poor capacity in adapting to
the changes in electricity consumption habits
which may occur due to the opening of new electricity markets, technological innovations, 
social and economic changes, to name a few. 
Hence it is strategically important to develop some new forecasting models 
which are more adaptive to ever-changing electricity consumption environment,
and the hybrid method proposed in this paper, designed for short-term forecasting for daily loads, 
represents a determined effort in this direction.

\begin{figure}[htbp]
\centering
\epsfig{file=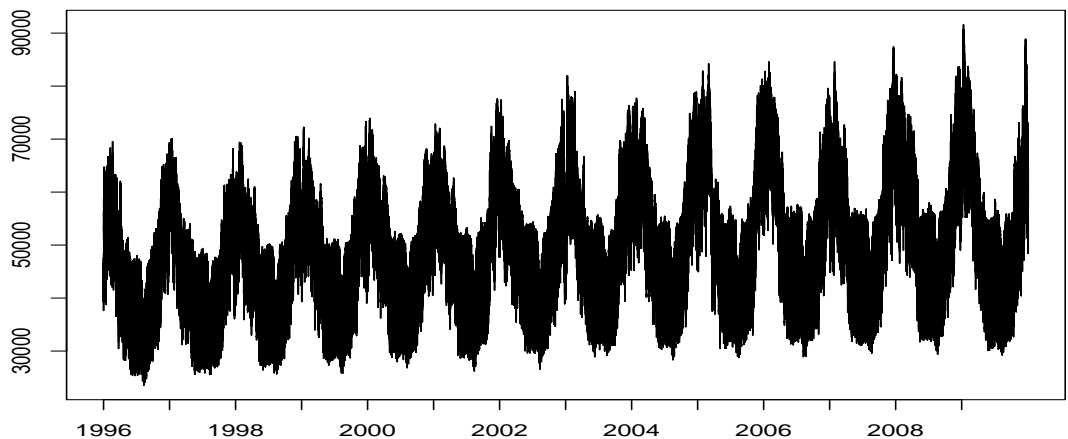, width=1\linewidth, height=3in} 
\caption{Electricity load from 1996 to 2009 in France.}
\label{fig:one}
\end{figure}

Electricity load exhibits interesting features at different levels.
Figure \ref{fig:one} displays the electricity load in France measured every half an hour from 1996 to 2009.
First of all, there is an overall increasing trend due to meteorological and economic factors.
In addition, an annual seasonal pattern repeats itself every year,
which can be explained by seasonal changes in temperature, day light duration and cloud cover.
\citet{Engl86} and \citet{Tayl02} discussed the impact of meteorological factors on the electricity load,
and singled out the temperature as being the most important due to the large demand of electrical heating in cold weather.
Further studies on the meteorological effect include \citet{Tayl08}, 
where seasonal patterns of electricity loads over 10 European countries were reported.
Also, there exist daily patterns which, unfortunately, do not show off due to the large scale of Figure \ref{fig:one}, 
attributed to varying demands for electricity in the different periods within a day.
Figure \ref{fig:four} below provides an example of such daily patterns.

Based on the above observations, we propose to model the electricity loads at
two different levels using different methods, hence the name \emph{hybrid} approach.
First assuming that the long-term trends do not vary greatly within a week, 
we extract those trends from weekly average loads using a generalised additive model,
where temperature and other meteorological factors are included as additional explanatory variables.
After removing the long term trend component from the data, 
we view the daily loads as curves and model the dynamic dependence 
among the electricity loads of successive days via curve linear regression. 
For this, a new dimension-reduction technique based on a singular value
decomposition in Hilbert space is proposed,
which reduces the regression with a curve response and a curve regressor 
to several ordinary (i.e. scalar) linear regression models.
Regarding the daily loads as curves, our approach takes advantage of the
continuity of the consumption curves in statistical modelling,
as well as embedding some nonstationary features (such as daily patterns)
into a stationary framework in a functional space.

When applied to electricity load forecasting, the proposed method is shown to provide more accurate predictions
than conventional methods such as those based on seasonal ARIMA models or exponential smoothing.
Although the operational model at EDF provides predictions of better accuracy than our method,
the latter is considerably simpler and does not make use of the full subject knowledge
that has been accumulated over more than 20 years at the EDF, which is not available in the public domain. 
Hence our approach is more adaptive to the changing electricity consumption environment 
while retaining a competitive prediction capacity,
and can be adopted as a generic tool applicable to a wide range of problems
including the electricity load forecasting in countries other than France. 
Furthermore it has the potential to serve as a building block for constructing a more
effective operational model when incorporating the full EDF subject knowledge.

There is a growing body of literature devoted to electricity load forecasting models.
Focusing on the main interest of this paper, we list below the recent papers on short-term load forecasting; 
see \citet{Bunn85} for a more comprehensive overview.
In the category of parametric approaches, \citet{Rama97} proposed linear regression models 
with autoregressive errors for each hour of a day.
Univariate methods such as those based on SARIMA models or 
exponential smoothing can be found in \citet{Hynd02}, \citet{Tayl06} and \citet{taylor2010}, 
and those based on state-space models in \citet{Dord08} and \citet{Dord11}.     
Among the nonparametric and semiparametric methods,
\citet{Engl86} proposed to include the temperature effect in the load modelling, 
and \citet{Harv93} proposed a time-varying spline model that captured both the temperature effect 
and the seasonal patterns in a semi-parametric way. 
Generalised additive models for electricity loads were studied in \citet{Pier11} and \citet{Hynd11}, 
where the semi-parametric approaches were shown to be well-adapted to non-linear behaviours of the electricity load signal. 
In \citet{Anto06}, a forecasting model based on functional data analysis was proposed
which treated the daily electricity loads as curves, and the approach has been further developed in \citet{cugliari2011}.
\citet{Cott03} proposed a Bayesian autoregressive model for short-term forecasts,
where the meteorological effects were estimated as non-linear using semi-parametric regression methods. 
They obtained good forecasting results with New South Wales dataset.
 
The rest of the paper is organised as follows. 
In Section \ref{sec:week}, we present the modelling of weekly average loads using a generalised additive model.
Then Section \ref{sec:day} discusses the modelling of the dependence structure between daily loads
in a curve linear regression framework.
We conduct a comparison study in Section \ref{sec:predict2009}, where our new method as well as other competitors 
are applied to predict the French daily loads in 2009.
Section \ref{sec:conclusion} contains some conclusive remarks.  
All the proofs are relegated to a supplementary document. 
 
\section{Modelling weekly averages}
\label{sec:week}

\begin{figure}[htbp]
\centering
\epsfig{file=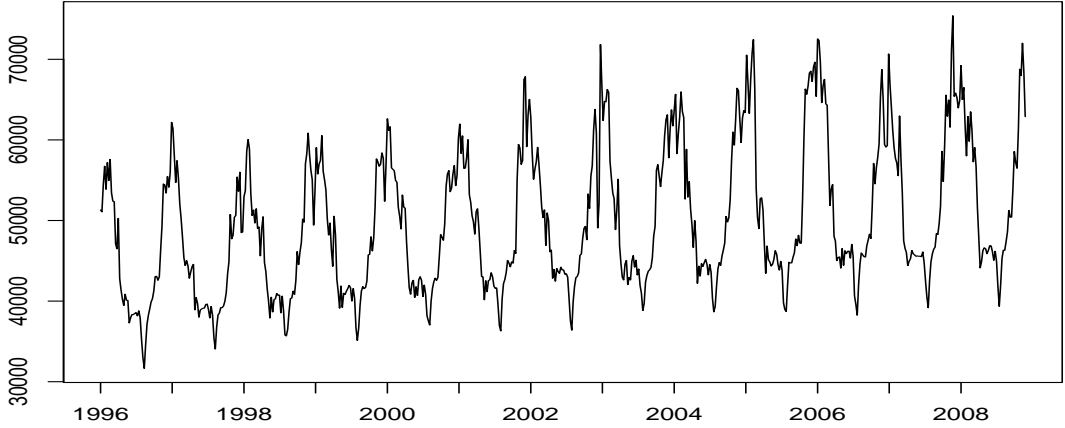, width=1\linewidth, height=3in} 
\caption{Weekly average electricity load in France from 1996 to 2008.}
\label{fig:two}
\end{figure}

Assuming that the overall trend and seasonality do not vary greatly within a week,
we propose to model the long-term trends with the \emph{weekly} averages, 
i.e. we treat the trend and seasonal component as being constant within each week.
In this manner, we lose little from the gradual changes of the trends within each week, 
while preserving the dependence structure across the electricity loads of different days.
The weekly averages of the EDF loads from 1996 to 2008 are plotted in Figure \ref{fig:two}.
In the literature, it has been noted that some meteorological factors, such as temperature and cloud cover, 
have a significant impact on the electricity consumption patterns.
While there are other de-trending techniques that have been proposed for removing long-term trends and seasonal cycles,
we fit the weekly averages using a generalised additive model (GAM)
for its ability to model implicit non-linear relationships between response and explanatory variables
without suffering from the so-called ``curse of dimensionality''; 
see \citet{hastie1990} and \citet{wood2006} for further details on the GAM,
and \citet{pierrot2009}, \citet{Pier11} and \citet{Hynd11} for its application in electricity load modelling.
Denoting the time index representing each week by $t$, 
the explanatory variables considered in fitting the weekly average load process $L_t$ are as follows:
$O_t$ is the weekly median of the \emph{offset} (a temporal variable determined by the experts at EDF 
to represent the seasonal trend in the data, taking values -3, -2, -1 and
0 to denote different winter holidays, 1 to denote spring, 2--6 to denote summer and summer holidays, and 7 to denote autumn),
$T_t$ is the weekly average of the temperature, $C_t$ is the weekly
average of the cloud cover,
and $I_t$ is the weekly index ranging from 1 to 53.

Our first attempt at taking into account the meteorological effects as well as the temporal trend 
is summarised in the following GAM with the Gaussian link function
\begin{eqnarray}
L_t = f_1(t) + f_2(O_t) + f_3(L_{t-1}) + f_4(T_t) + f_5(T_{t-1}) + f_6(C_t),
\label{eq:trend:one}
\end{eqnarray}
where each $f_j$ is a smooth function of the corresponding covariate 
with thin plate regression splines as a smoothing basis.
We use the R package \emph{mgcv} introduced in \citet{wood2006},
where each smooth function $f_j$ is estimated by penalised regression splines.
In this implementation, the amount of penalisation is calibrated according to the generalised cross-validation (GCV) score,
see \citet{Wood04} and \citet{Wood11} for details.

\begin{figure}[htbp]
\psfrag{GAM (1)}{\scriptsize{GAM (1)}}
\psfrag{GAM (2)}{\scriptsize{GAM (3)}}
\centering
\epsfig{file=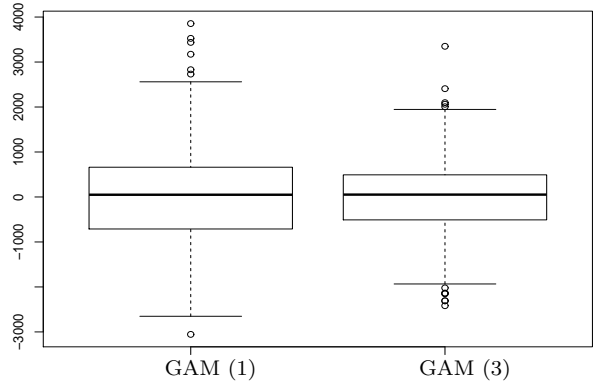, width=.55\linewidth, height=2.5in} 
\caption{Boxplots of the residuals from fitting the weekly average load between 1996 and 2008
using the model (\ref{eq:trend:one}) (left) and the model (\ref{eq:trend:two}) (right).}
\label{fig:boxplot}
\end{figure}

\begin{figure}[htbp]
\psfrag{t}{$\scriptstyle{t}$}
\psfrag{Ot}{$\scriptstyle{O_t}$}
\psfrag{L(t-1)}{$\scriptstyle{L_{t-1}}$}
\psfrag{Tt}{$\scriptstyle{T_t}$}
\psfrag{T(t-1)}{$\scriptstyle{T_{t-1}}$}
\psfrag{Ct}{$\scriptstyle{C_t}$}
\psfrag{f1(t)}{$\scriptstyle{f_1(t)}$}
\psfrag{f2(Ot)}{$\scriptstyle{f_2(O_t)}$}
\psfrag{f3(L(t-1))}{$\scriptstyle{f_3(L_{t-1})}$}
\psfrag{f4(Tt)}{$\scriptstyle{f_4(T_t)}$}
\psfrag{f5(T(t-1))}{$\scriptstyle{f_5(T_{t-1})}$}
\psfrag{f6(Ct)}{$\scriptstyle{f_6(C_t)}$}
\begin{center}
\epsfig{file=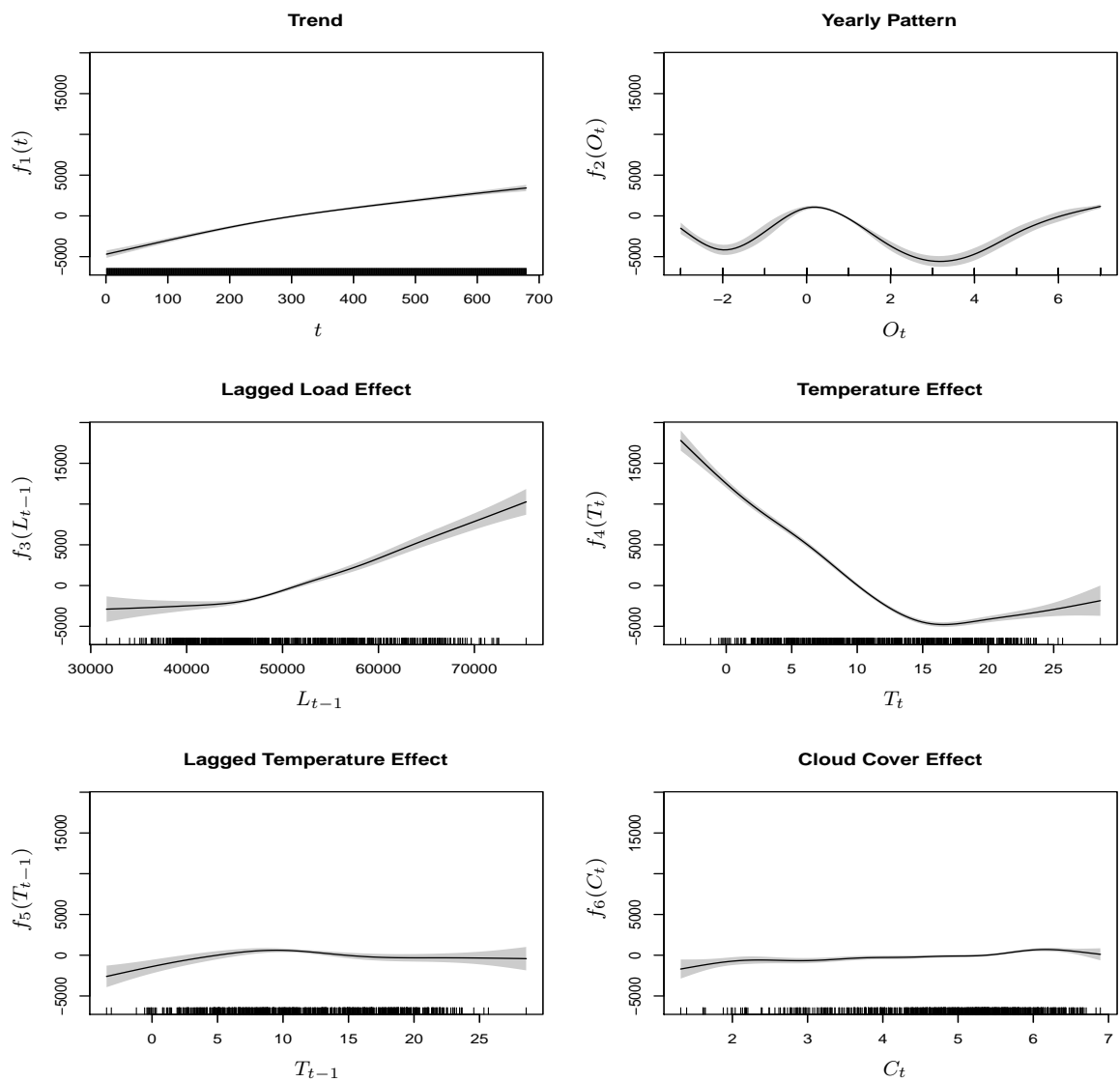, width=1\linewidth, height=6in}  
\end{center}
\caption{Estimated $f_1, \ldots, f_6$ from model (\ref{eq:trend:one}); 
shaded regions represent the confidence bands.}
\label{fig:gam:one}
\end{figure}

We note that the basis used to estimate $f_1$ has knots at each first week of September,
which are imposed to model the time-varying trend in the electricity load at the yearly level.
The boxplot of the residuals from fitting the above GAM to the weekly average load between 1996 and 2008
is provided in Figure \ref{fig:boxplot}, 
and the estimated curves for
 $f_1, \ldots, f_6$ in (\ref{eq:trend:one}) are plotted in Figure \ref{fig:gam:one},
with shaded area representing the twice standard error bands below and above the estimate.
The fitted curve explains 98.7\% of the data, and 
the mean absolute percentage error (MAPE) and the root mean square error (RMSE)
from the estimated curve are 1.63\% and 1014MW, respectively.
The two error measures, MAPE and RMSE, are defined as
\begin{eqnarray}
\label{def:errors} 
{\rm MAPE} = \frac{1}{T} \sum_{t=1}^T \left\vert \frac{\wh L_t - L_t}{L_t} \right\vert \mbox{\quad and \quad}
{\rm RMSE} = \left\{ \frac{1}{T} \sum_{t=1}^T (\wh L_t - L_t)^2 \right\}^{1/2},
\end{eqnarray}
where $\wh L_t$ denotes the estimated (or predicted) load in the week $t$. 

We state below some observations based on the estimated functions in Figure \ref{fig:gam:one}.
The top left panel shows that the electricity load increases over time $t$, and that the trend is almost linear.
The top right panel shows clearly the presence of the  seasonality 
as the load is lower during holidays and in summer than that in winter.
As for the lagged load effect, $L_t$ increases with respect to its lagged
value $L_{t-1}$ (the second left panel)
and the rate of increase is greater when $L_{t-1} > 5\times 10^4$ approximately,
which implies that the value $5\times 10^4$ may be regarded as a ``threshold'' 
acting on the impact of $L_{t-1}$ on $L_t$.
Since the increase in the usage of electricity is closely related to the climate, 
which in turn is linked to the time of the year, we may include the joint effect of $L_{t-1}$ and $I_t$ in the model
to accommodate the dependence between those two variables.
Also, the impact of temperature is significant (the second right panel).
The low temperatures lead to high electricity consumptions due to electrical heating,
resulting the initial sharp decrease in $\wh f_4$. 
Then as the temperature increases from about 17$^\circ$C upwards, $\wh f_4$ also increases slowly, 
which can be accounted by the use of cooling system in hot weather.
As the meteorological changes within a year is closely related to the time index, 
we may include the joint effect of the variables $T_t$ and $I_t$ in the model.
The bottom panels show that, although not as prominent as other terms, 
the lagged temperature and the cloud cover do have an impact on the weekly average load 
at large values of $T_{t-1}$ and $C_t$. 
The effect of cloud cover is significantly different from 0 for large values of $C_t$,
as heavy cloud cover induces the increasing use of lighting (the bottom right panel).
We note that the estimated effect of the low cloud cover may be an artifact:
there are only few observations available for low cloud cover and  
thus the variance of the fitted curve at such small values of $C_t$ is large.

\begin{figure}[htbp]
\psfrag{L[t]}{\begin{sideways}$\scriptstyle{L_t}$\end{sideways}}
\psfrag{I[t]}{$\scriptstyle{I_t}$}
\psfrag{L[t - 1]}{$\scriptstyle{L_{t-1}}$}
\psfrag{T[t]}{$\scriptstyle{T_t}$}
\begin{tabular}{cc}
\epsfig{file=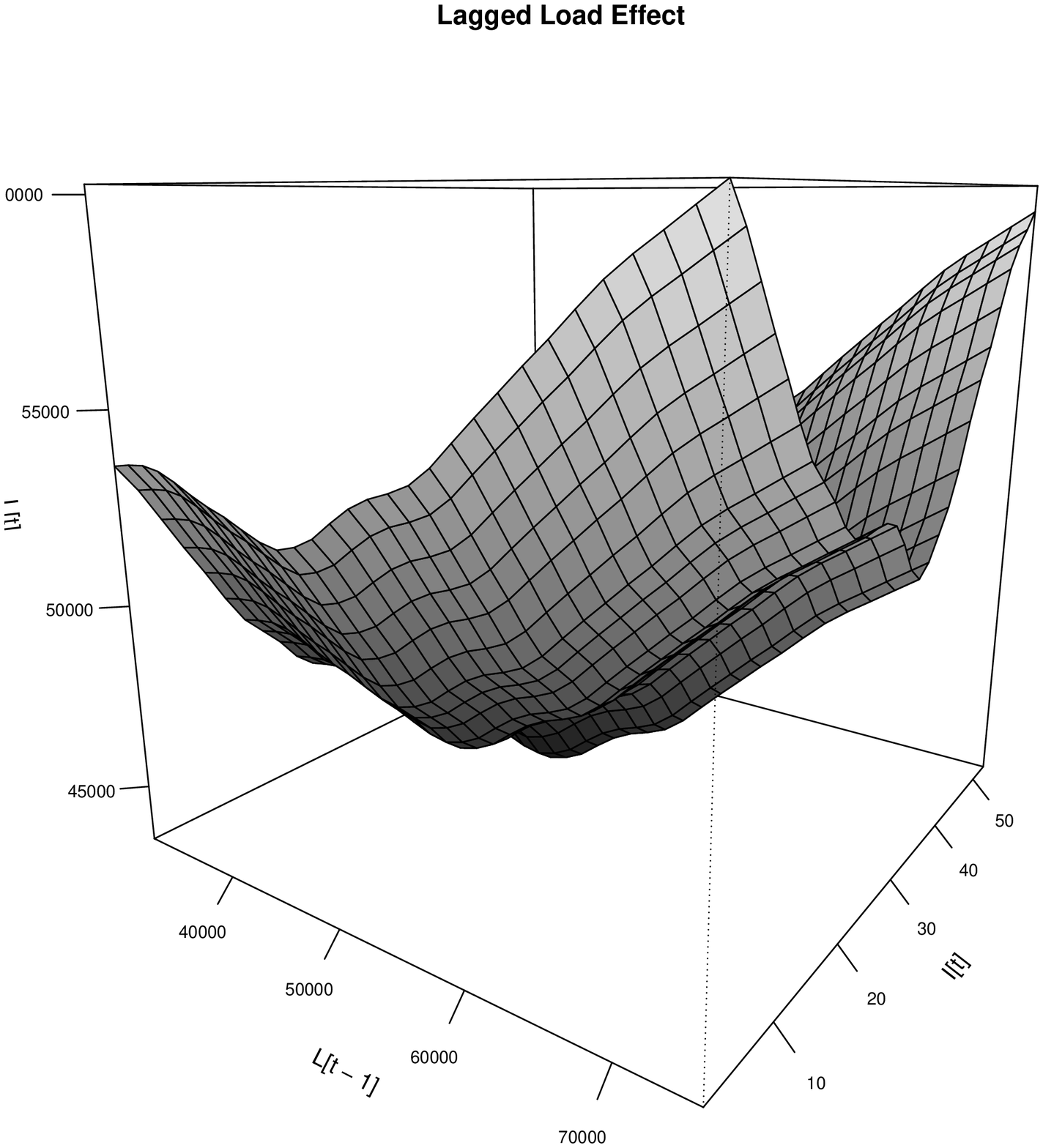, width=.5\linewidth, height=3in} &
\epsfig{file=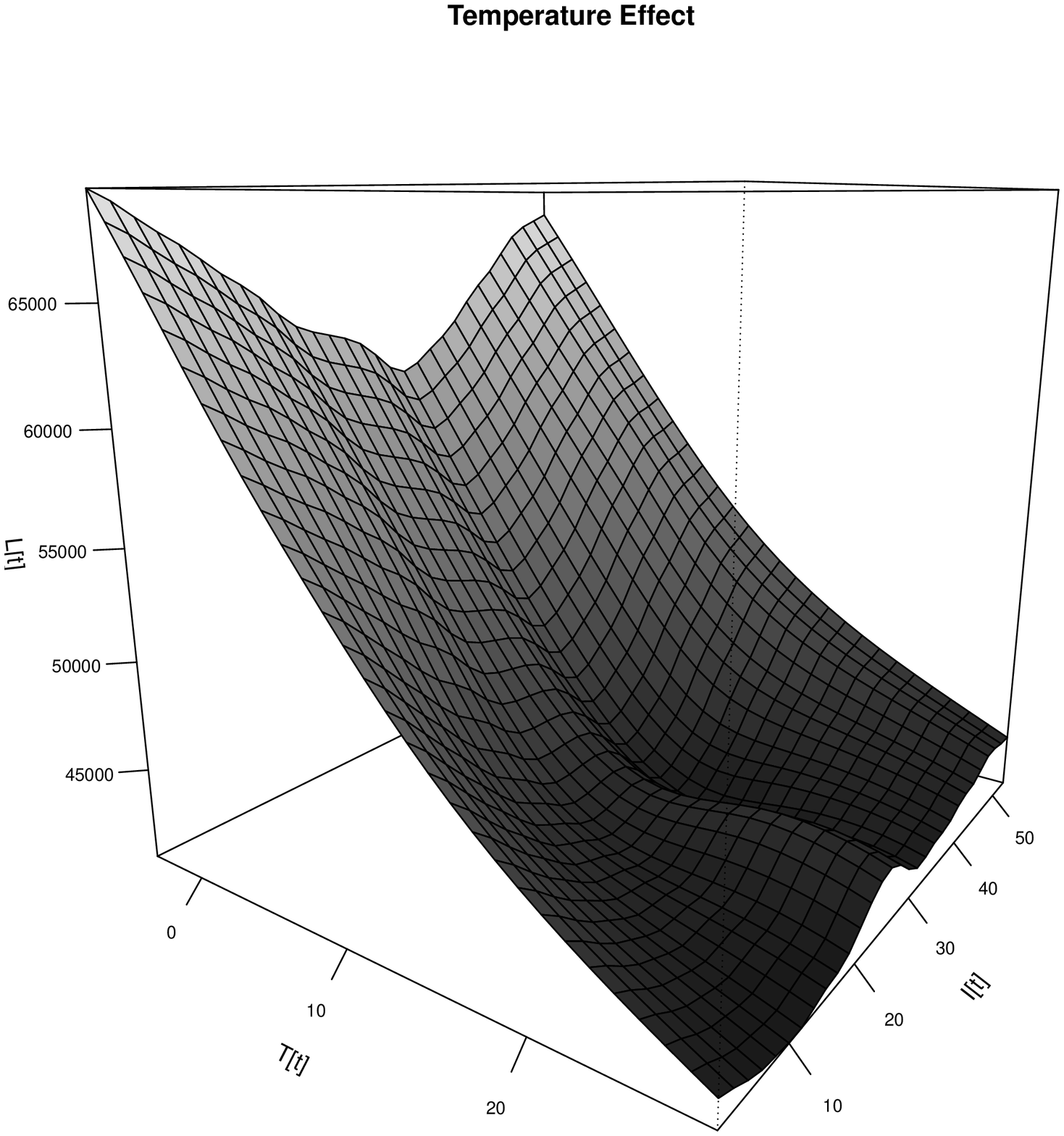, width=.5\linewidth, height=3in}
\end{tabular}
\caption{Estimated $f_3$ (left) and $f_4$ (right) from model (\ref{eq:trend:two}).}
\label{fig:gam:two}
\end{figure}

Based on the above observations, we propose another model
\begin{eqnarray}
L_t = f_1(t) + f_2(O_t) + f_3(L_{t-1}, I_t) + f_4(T_t, I_t) + f_5(T_{t-1}, I_t) + f_6(C_t, I_t),
\label{eq:trend:two}
\end{eqnarray}
where $f_3, \ldots, f_6$ include the weekly index $I_t$ as a covariate.
To study the bivariate effects, the estimated $f_3$ and $f_4$ are plotted in Figure \ref{fig:gam:two}.
The impact of the lagged load $L_{t-1}$ on the load $L_t$ is similar as previously described in the sense that, 
the rate of increase of $L_t$ changes when $L_{t-1}$ is greater than a threshold value.
However, we also note that the relationship between $L_{t-1}$ and $L_t$ varies throughout a year 
with the weekly index $I_t$, and that the impact of $L_{t-1}$ is far stronger in winter than in summer. 
As for the effect of temperature, there is a smooth transition observable throughout a year 
from the winter heating effect to the summer cooling effect.

With the new model, there is an increase in the percentage of the data explained (99.2\%),
and both the MAPE (1.28\%) and the RMSE (801MW) of the fitted trend have decreased.
Further, the GCV score indicates that the new model is favourable 
($8.4 \times 10^5$) to the previous one ($1.2 \times 10^6$).
Also, when comparing the forecasts from the two models for the weekly average loads of 2009, (\ref{eq:trend:two}) 
performed considerably better (MAPE 1.72\%, RMSE 1250MW) than the model (\ref{eq:trend:one}) (MAPE 2.15\%, RMSE 1532MW).
We note that the superior performance of the model (\ref{eq:trend:two}) at the weekly level 
carries over to that at the daily electricity load forecasting;
when applied to forecast the daily loads in 2009, 
the MAPE and RMSE from the model (\ref{eq:trend:two}) were 1.35\% and 869MW respectively,
whereas the model (\ref{eq:trend:one}) led to 1.41\% and 901MW
(see Section \ref{sec:predict2009} for full details of the forecasting procedure).
From these observations and also from the residual boxplots in Figure \ref{fig:boxplot},
we choose model (\ref{eq:trend:two}) over model (\ref{eq:trend:one}).

\section{Regression of daily load curves}
\label{sec:day}

Once the long-term trend is fitted as in Section \ref{sec:week} and removed, 
we regard the residuals on the $i$-th day as a curve $Y_i(\cdot)$ defined on the index set $\cI_1$,
and model the dependency among the daily loads via curve linear regression as 
\begin{eqnarray}
\label{curve:lm}
Y_i(u)=\int_{\cI_2}X_i(v)\beta(u, v)dv + \vep_i(u) \mbox{ \quad for \quad } u \in \cI_1,
\end{eqnarray}
where $X_i(\cdot)$ can be, for example, the residual curve on the ($i-$1)-th day
(i.e. $Y_{i-1}(\cdot)$),
or the curve joining $Y_{i-1}(\cdot)$ and the temperature curve on the $i$-th day.
Therefore the index set of $X_i(\cdot)$, say $\cI_2$, may be different from $\cI_1$.
In (\ref{curve:lm}), $\beta$ is a regression coefficient function defined on $\cI_1 \times \cI_2$,
and $\vep_i(\cdot)$ is noise with mean 0.

Linear regression with curves as both response and regressor, has been
studied by \citet{dalzell1991}, \citet{hmw2000}, and \citet{cmw2004} and \citet{ymw2005} among others. 
The conventional approach is to apply the Karhunen-Lo\`eve decomposition to both $Y_i(\cdot)$ and $X_i(\cdot)$,
and then to fit a regression model using the finite number of terms obtained from such decompositions. 
The Karhunen-Lo\`eve decomposition has featured predominantly in functional data analysis; 
see also \citet{fanzhang1998} and \citet{horowitz2007}.
This approach is identical to the dimension reduction based on principal component analysis in multivariate analysis.
Since the principal components do not necessarily represent the directions in which
$X_i(\cdot)$ and $Y_i(\cdot)$ are most correlated, 
we present below a novel approach where the singular value decomposition (SVD) is applied 
to single out the directions upon which the projections of $Y_i(\cdot)$ are most correlated with $X_i(\cdot)$. 
Our method is closely related to the canonical correlation analysis yet 
we focus on regressing $Y_i(\cdot)$ on $X_i(\cdot)$, and thus
$Y_i(\cdot)$ and $X_i(\cdot)$ are not treated on an equal footing,
which is different from, and much simpler than, the canonical correlation analysis.
The literature on functional canonical correlation analysis includes
\citet{hannan1961}, \citet{silverman1996}, \citet{he2003}, \citet{cupidon2008}, \citet{eubank2008} and \citet{yang2011}.

\subsection{Curve linear regression via dimension reduction}
\label{sec:curve}

Let $\{Y_i(\cdot) , X_i(\cdot) \}, \; i=1, \ldots, n$, be a random sample where
$Y_i(\cdot) \in \cL_2(\cI_1)$, $X_i(\cdot) \in \cL_2(\cI_2)$, 
and let $\cI_1$ and $\cI_2$ be two compact subsets of $\R$. 
We denote by $\cL_2(\cI)$ the Hilbert space consisting of all the square integrable curves defined on the set $\cI$, 
which is equipped with the inner product 
$\inner{f}{g} = \int_\cI f(u)g(u) du$ for any $f, g \in \cL_2(\cI)$.
We assume that $\E\{Y_i(u)\}=0$ for all $u \in \cI_1$ and $\E\{X_i(v)\}=0$ for all $v \in \cI_2$,
and denote the covariance function between $Y_i(\cdot)$ and $X_i(\cdot)$ by $\Sigma(u ,v) = \cov\{Y_i(u),\,X_i(v)\}$.
Under the assumption
\begin{eqnarray}
\label{l2n}
\int_{\cI_1}\E\{Y_i(u)^2\}du + \int_{\cI_2}\E\{X_i(v)^2\}dv < \infty,
\end{eqnarray}
$\Sigma$ defines the following two bounded operators between $\cL_2(\cI_1)$ and $\cL_2(\cI_2)$: 
\begin{eqnarray*}
f_1(u) \to \int_{\cI_1} \Sigma(u,v) f_1(u) du \in \cL_2(\cI_2), \qquad
f_2(v) \to \int_{\cI_2} \Sigma(u,v) f_2(v) dv \in \cL_2(\cI_1)
\end{eqnarray*}
for any $f_i \in \cL_2(\cI_i)$. 
Based on the SVD, there exists a triple sequence $\{(\varphi_j, \psi_j, \lam_j), \; j=1, 2, \dots\}$ for which
\begin{eqnarray}
\label{sigma}
\Sigma(u, v) = \sum_{j=1}^\infty \sqrt{\lam_j}\;\varphi_j(u)\,\psi_j(v),
\end{eqnarray}
where $\{\varphi_j\}$ is an orthonormal basis of $\cL_2(\cI_1)$, 
$\{\psi_j\}$ is an orthonormal basis of $ \cL_2(\cI_2)$, and $\{\lam_j\}$ are ordered such that
\begin{eqnarray}
\label{lam:order}
\lam_1 \ge \lam_2 \ge \cdots \ge 0.
\end{eqnarray}
Further, it holds that for $u\in \cI_1, \; v\in \cI_2$ and $j=1, 2, \ldots$,
\begin{eqnarray}
\label{m:eig}
\int_{\cI_1} M_1(u, z)\,\varphi_j(z)\,dz = \lam_j\,\varphi_j(u), \qquad
\int_{\cI_2} M_2(v, z)\,\psi_j(z)\,dz = \lam_j\,\psi_j(v), 
\end{eqnarray}
where $M_i$ is a non-negative operator defined on $\cL_2(\cI_i)$ as
\begin{eqnarray*}
M_1(u, u') = \int_{\cI_2}\Sigma(u, z)\,\Sigma(u', z)\,dz, \qquad
M_2(v, v') = \int_{\cI_1}\Sigma(z, v)\,\Sigma(z, v')\,dz.
\end{eqnarray*}
It is clear from (\ref{m:eig}) that $\lam_j$ is the $j$-th largest eigenvalue of $M_1$ and $M_2$ with 
$\varphi_j$ and $\psi_j$ as the corresponding eigenfunctions, respectively.
Since $\{\varphi_j\}$ and $\{\psi_j\}$ are the orthonormal basis of $\cL_2(\cI_1)$ and $\cL_2(\cI_2)$, we may write
\begin{eqnarray}
\label{yx:eig}
Y_i(u) = \sum_{j=1}^\infty \xi_{ij} \varphi_j(u),
\qquad 
X_i(v) = \sum_{j=1}^\infty \eta_{ij} \psi_j(v),
\end{eqnarray}
where $\xi_{ij}$ and $\eta_{ij}$ are random variables defined as
\begin{eqnarray} \label{q3} 
\xi_{ij}  = \int_{\cI_1} Y_i(u) \varphi_j(u) du, \qquad
\eta_{ij}  = \int_{\cI_2} X_i(v) \psi_j(v) dv.
\end{eqnarray}
It follows from (\ref{sigma}) that
\begin{eqnarray}
\label{xi:eta}
\cov(\xi_{ij}, \, \eta_{ik}) = \E(\xi_{ij}\eta_{ik}) = 
\left\{\begin{array}{ll}
\sqrt{\lam_j} \qquad & \mbox{for \ } j=k, \\
0 \qquad & \mbox{for \ } j \ne k.
\end{array}\right.
\end{eqnarray}
We refer to \citet{smithies1937} for further details on the SVD in a Hilbert space.

Now, we are ready to introduce the notion of the correlation dimension between the two curves. 
See \citet{hallvial2006} and \citet{qiwei2010} for the definitions of curve dimensionality in different contexts.
\begin{defn}
\label{def:one}
The correlation between curves $Y_i(\cdot)$ and $X_i(\cdot)$ is $r$-dimensional if $\lam_r>0$ and $\lam_{r+1}=0$ in (\ref{lam:order}).
\end{defn} 
When the correlation between $Y_i(\cdot)$ and $X_i(\cdot)$ is $r$-dimensional, 
it follows from (\ref{xi:eta}) that  
$\cov\{\xi_{ij}, \, X_i(v)\}=0$ for all $j>r$ and $v\in\cI_2$.
Moreover, the curve linear regression model (\ref{curve:lm}) admits an equivalent representation with $r$ 
(scalar) linear regression models; see Theorem \ref{thm:one} below.
Before presenting the theorem, we further assume that
the regression coefficient $\beta(u, v)$ is in the Hilbert space $\cL_2(\cI_1\times\cI_2)$,
and that $\vep_i(\cdot)$ are i.i.d. with $\E\{\vep_i(u)\}=0$ and $\E\{X_i(v)\vep_j(u)\}=0$ 
for any $u \in \cI_1, \; v\in \cI_2$ and $i, j \ge 1$.
\begin{thm}
\label{thm:one}
Let the linear correlation between $Y_i(\cdot)$ and $X_i(\cdot)$ be $r$-dimensional. 
Then the curve regression (\ref{curve:lm}) may be represented equivalently by
\begin{eqnarray}
\label{multiple:lm}
\begin{array}{ll}
\xi_{ij} = \sum_{k=1}^\infty \beta_{jk}\eta_{ik} + \vep_{ij} & \qquad \mbox{for \ } j=1, \ldots, r, \\
\xi_{ij} = \vep_{ij} & \qquad \mbox{for \ } j=r+1, r+2, \ldots,
\end{array}
\end{eqnarray}
where
$\vep_{ij} = \int_{\cI_1} \varphi_j(u)\vep_i(u) du$, and
$\beta_{jk} = \int_{\cI_1\times\cI_2} \varphi_j(u)\psi_k(v)\beta(u, v) dudv$.
\end{thm} 

The proof of Theorem \ref{thm:one} can be found in the supplementary document. 
Some remarks are listed in order.
\begin{itemize}
\item[(a)] 
For each $j=1, \ldots, r$, we may apply model selection criteria such as the AIC,
to select the variables to be included in the first linear regression model of (\ref{multiple:lm})
among $\{\eta_{ik}, \, k \ge 1\}$, noting $\var(\eta_{ik}) \to 0$ as $k\to\infty$; see (\ref{l2n}) and (\ref{yx:eig}).
We also note that $\{\varphi_j(u)\psi_k(v)\}_{j,k}$ form an orthonormal basis of $\cL_2(\cI_1\times\cI_2)$. 
Since $\beta(u, v)\in \cL_2(\cI_1 \times \cI_2)$, it holds that
$\sum_{j=1}^\infty\sum_{k=1}^\infty\beta_{jk}^2 = \int_{\cI_1\times\cI_2}\beta(u, v)^2dudv < \infty$.

\item[(b)] 
In fact, Theorem \ref{thm:one} holds for any valid expansion of $X_i(v)$ as $X_i(v) = \sum_k\eta_{ik}\psi_k(v)$,
provided $\{\xi_{ij}\}$ are obtained from the SVD. 
For example, we may use the Karhunen-Lo\`eve decomposition of $X_i(\cdot)$.
Then resulting $\eta_{ik}$ is the projection of $X_i(\cdot)$ on the $k$-th principal direction, 
and those $\{\eta_{ik}\}$ are uncorrelated with each other. 

\item[(c)]
Let $X_i(\cdot)$ be of finite dimension in the sense that its Karhunen-Lo\`eve decomposition has $q$ terms only as
$X_i(v)=\sum_{k=1}^{q} \zeta_{ik}\gam_k(v)$,
where $q (\ge r)$ is a finite integer, 
$\{\gam_k(\cdot)\}_{k=1}^q$ are $q$ orthonormal functions in $\cL_2(\cI_2)$,
and $\zeta_{i1}, \ldots, \zeta_{iq}$ are uncorrelated with $\var(\zeta_{ik})>0$ for all $k=1, \ldots, q$.
Without loss of generality, we may assume that $\var(\zeta_{ik})=1$, 
which can be achieved by replacing $X_i(v)$ with its linear transformation
$\int_{\cI_2}\Gamma(v, w)X_i(w)dw$, where
$\Gamma(v, w)=\sum_{k=1}^{q} \gam_k(v)\gam_k(w) \big/ \sqrt{\var(\zeta_{ik})}$.
Then for such $X_i(\cdot)$, the second equation in (\ref{yx:eig}) is reduced to 
$X_i(v) = \sum_{k=1}^{q} \eta_{ik}\psi_k(v)$
with $\{\eta_{ik}\}$ satisfying $\var(\eta_{ik})=1$ and $\cov(\eta_{ik}, \eta_{il}) = 0$ for any $k \ne l$.
This, together with (\ref{xi:eta}) and (\ref{multiple:lm}), implies that $\beta_{jk}=0$ in (\ref{multiple:lm}) for all $j \ne k$. 
Hence (\ref{multiple:lm}) is reduced to
\begin{eqnarray}
\label{simple:lm}
\begin{array}{ll}
\xi_{ij} = \beta_{jj}\eta_{ij} + \vep_{ij} & \qquad \mbox{for \ } j=1, \ldots, r,\\
\xi_{ij} = \vep_{ij} & \qquad \mbox{for \ } j=r+1, r+2, \ldots,
\end{array}
\end{eqnarray}
i.e. under the additional condition on the dimensionality of $X_i(\cdot)$, 
the curve regression (\ref{curve:lm}) is reduced to $r$ simple linear regression problems. 

\item[(d)]
We provide a recap of the above results in the context of vector regression.
Let $\by_i$ and $\bx_i$ be, respectively, $p\times 1$ and $q\times 1$ vectors.
Suppose that $\rk(\bSig_{yx})=r$, where $\bSig_{yx} = \cov(\by_i, \bx_i)$.
Then the multiple linear regression problem $\by_i = \bB\bx_i + \bvep_i$
may be reduced to the $r$ scalar linear regression problems:
\begin{eqnarray} \label{b15}
u_{ij} = \bv_i '\bbeta_j + \epsilon_{ij}, \qquad j=1, \cdots, r.
\end{eqnarray}
Here, $(u_{i1}, \cdots, u_{ip})' = \bU '\by_i$ and $\bv_i=(v_{i1}, \cdots, v_{iq})' = \bV '\bx_i$.
Also, $\bSig_{yx} = \bU\bLam\bV '$ is the SVD of $\bSig_{yx}$ with $\bU \bU '=\bI_p$, $\bV \bV '=\bI_q$ and
$\bLam$ is a $p \times q$ diagonal matrix with only the first $r (\le \min(p, q))$
main diagonal elements being nonzero.
If $\var(\bx_i) = \sigma^2 \bI_q$ is satisfied in addition, (\ref{b15}) reduces to $r$ simple regression models
$u_{ij} =  v_{ij}\beta_j + \vep_{ij}$ for $j=1, \cdots, r$.
\end{itemize}

\subsection{Estimation}
\label{sec:estimation}

We assume the availability of observed curves $\{Y_i(\cdot), \, X_i(\cdot)\}$ for $i=1, \cdots, n$.
Recalling $\Sigma(u, v) = \cov\{Y_i(u), X_i(v)\}$, let
\begin{eqnarray*}
\wh \Sigma(u, v) = \frac{1}{n} \sum_{i=1}^n \{Y_i(u) - \bar Y(u)\}\{X_i(v) - \bar X(v)\},
\end{eqnarray*}
where $\bar Y(u) = n^{-1} \sum_i Y_i(u)$ and $\bar X(v) = n^{-1} \sum_i X_i(v)$.
Performing the SVD on $\wh \Sigma(u,v)$, we obtain the estimators 
$(\wh \lam_j, \, \wh \varphi_j , \, \wh \psi_j)$ for 
$(\lam_j, \, \varphi_j,\, \psi_j)$ as defined in (\ref{sigma}).
Note that this SVD is effectively an eigenanalysis of the non-negative operator 
\begin{eqnarray} \label{hatm1}
\wh M_1(u, u') = \int_{\cI_2} \wh \Sigma(u, v) \wh\Sigma(u', v)dv,
\end{eqnarray}
which may be transformed into an eigenanalysis of a non-negative definite matrix.
Furthermore $\wh \varphi_j(\cdot)$ and $\wh \psi_j(\cdot)$ may be taken as linear
combinations of, respectively, the observed curves $Y_i(\cdot)$ and $X_i(\cdot)$.
See, for example, Section 2.2.2 of \citet{qiwei2010}.
Proposition~\ref{prop:lam} below presents the asymptotic properties for the estimators $\wh \lam_j$. 
Its proof is similar to that of Theorem~1 of \citet{qiwei2010} and is thus omitted.
\begin{prop}
\label{prop:lam} 
Suppose that $\{Y_i(\cdot), X_i(\cdot)\}$ is strictly stationary and $\psi$-mixing
with the mixing coefficients $\psi(k)$ satisfying the condition
$\sum_{k \ge 1} k\psi(k)^{1/2} < \infty$. 
Further, assume $\E\{\int_{\cI_1} Y_i(u)^2 du+ \int_{\cI_2} X_i(v)^2dv\}^2 < \infty$ and 
let $\lam_1 > \cdots > \lam_r >0 = \lam_{r+1} = \lam_{r+2} = \cdots$.
Then as $n \to \infty$, 
\begin{itemize}
\item[(i)] $\vert\wh \lam_k-\lam_k\vert = O_p(n^{-1/2})$ for $1\le k\le r$, and
\item[(ii)] $\vert\wh \lam_k\vert = O_p(n^{-1})$ for $k>r$.
\end{itemize}
\end{prop}
We refer to Section 2.6 of \citet{fanyao2003} for the further details on mixing conditions. 
The fast convergence for the zero-eigenvalues $\lam_j$ with $j>r$ is due to the quadratic form in (\ref{hatm1}), 
and the relevant discussion is provided in \citet{qiwei2010} and \citet{lamyao2011}. 
It follows from Proposition~\ref{prop:lam} that the ratios $\wh \lam_{j+1} / \wh \lam_j$ for $j<r$ 
are asymptotically bounded away from 0, and $\wh \lam_{r+1}/ \wh \lam_r \to 0$ in probability. 
This motivates the following ratio-based estimator. 
In \citet{lamyao2011}, a more elaborate investigation of this estimator can be found in a different context.
\begin{description}
\item[The ratio-based estimator for the correlation dimension $r$:] \hfill \\
$\wh r = \arg\min_{1\le j \le d} \wh \lam_{j+1} / \wh \lam_j$, 
where $d>r$ is a fixed and pre-specified integer.
\end{description}
One alternative is to use properly defined information criteria as in, e.g. \citet{hallin2007},
where a similar idea was adopted for high-dimensional time series analysis. To this end, we define 
\begin{eqnarray*}
IC_1(q) = \frac{1}{d^2}\sum_{k=q+1}^d \wh\lam_k + \tau_1q \cdot g(n), \mbox{ \ and \ } 
IC_2(q) = \log\left(c_* + \frac{1}{d^2}\sum_{k=q+1}^d \wh\lam_k\right) + \tau_2q \cdot g(n),
\end{eqnarray*}
where $c_*, \tau_1, \tau_2>0$ are constants, $d>r$ is a pre-specified integer, and $g(n)>0$ satisfies
\begin{eqnarray}
n \cdot g(n)\to\infty \quad \mbox{and} \quad g(n) \to 0, \quad {\rm as} \quad n \to \infty.
\label{eq:gdt}
\end{eqnarray}
Theorem~\ref{prop:one} below shows that
$\wh r \equiv \arg\min_{0 \le q < d} IC_i(q)$ is a consistent estimator of $r$ for both $i=1, 2$.
The proof is given in the supplementary document.
\begin{thm}
\label{prop:one}
Let the conditions of Proposition~\ref{prop:lam} hold and both $r$ and $d$ be fixed as $n \to \infty$.
Then, for both $i=1,2$, we have $P\{IC_i(r) < IC_i(q)\} \to 1$ for any $0 \le q < d$ and $q \ne r$.
\end{thm}
\vspace{3mm}
The choice of $c_*$ is not critical as it is introduced to ensure that the term inside the logarithm is positive.
The proof of Theorem~\ref{prop:one} indicates that the consistency holds for
any constants $\tau_1$ and $\tau_2$. 
However, they affect the finite sample performance of the method and therefore
in practice, the choice of the tuning parameters and $\tau_1$ and $\tau_2$ and 
the penalty function $g(n)$ requires more care.
In our data analysis, we set $g(n)=n^{-1/2}$ and 
elaborate the choice of $\tau_i$ using the following majority voting scheme.

We start with two values $\tau_*$ and $\tau^*$ such that
$IC_i(q)$ is minimised at $q=d$ for any $\tau_i \le \tau_*$, and at $q=0$ for any $\tau_i \ge \tau^*$.
Over the interval $[\tau_*,\, \tau^*]$, the function $h(\tau)\equiv\arg\min_qIC_{i}(q)$ is non-increasing in $\tau$.
Then, assigning a grid of values from $[\tau_*,\, \tau^*]$ as $\tau_i$,
we look for the $q$ that is returned over the longest interval of $\tau_i$ within $[\tau_*,\, \tau^*]$,
and set such $q$ as the estimate of $r$.
Figure \ref{fig:ex:ic} below shows an example of applying $IC_2(q)$ for the selection of $r$,
where $IC_2(q)$ is computed over $q=1, \ldots, 20$ for 100 different values of $\tau_2$.
In this example, $q=4$ was returned most frequently as the minimiser of $IC_2(q)$.
We have further conducted a simulation study to check whether the proposed scheme 
worked well on simulated datasets of varying dimensionalities,
and the results have confirmed its good performance over a range of $r$.

\subsection{An illustration}
\label{sec:illustration}

\begin{figure}[htbp]
\centering
\epsfig{file=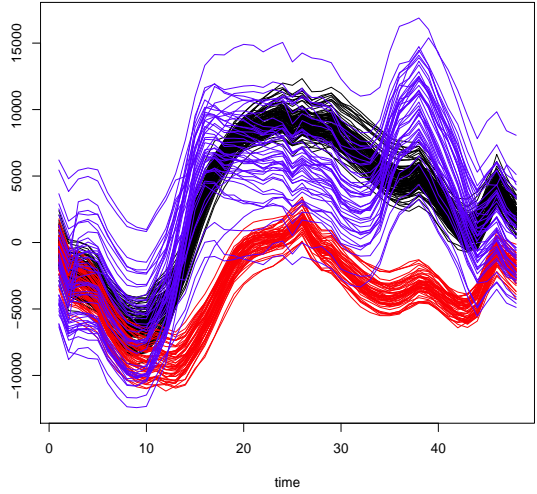, width=.5\linewidth, height=3in} 
\caption{De-trended daily curves for Tuesdays in July (black), Saturdays in July (red) and 
Tuesdays in December (blue) between 1996 and 2008.}
\label{fig:four}
\end{figure}

We illustrate the hybrid approach by predicting the load curve on 2 April 2009, which is denoted by $Z(\cdot)$. 
Unfortunately, even after removing the long-term trend estimated in Section~\ref{sec:week},
there exist some systematic discrepancies among the profiles of daily load curves over
different days in a week and different months in a year.
Figure \ref{fig:four} shows that, while the daily loads on Tuesdays in July are similar to each other,
they are distinctively different from those on Saturdays in July, and also
from those on Tuesdays in December. 
Those profile differences are reflected predominantly in the locations and magnitudes of daily peaks. 
Typically in France, daily peaks occur at noon in summer and in the evening in winter, 
due to the economic cycle as well as the usage of electrical heating and lighting.
Hence, the daily curves and presumably their dynamic structure vary over different days within a week, 
and also over different months in a year; 
further elaboration on those features is provided in Section~\ref{sec:predict2009} below.

\begin{figure}[htbp]
\psfrag{Class A}{$\scriptstyle X^{\mbox{\tiny L}}_i$}
\psfrag{Class C}{$\scriptstyle X^{\mbox{\tiny T}}_i$}
\psfrag{Class B}{$\scriptstyle Y_i$}
\begin{center}
\epsfig{file=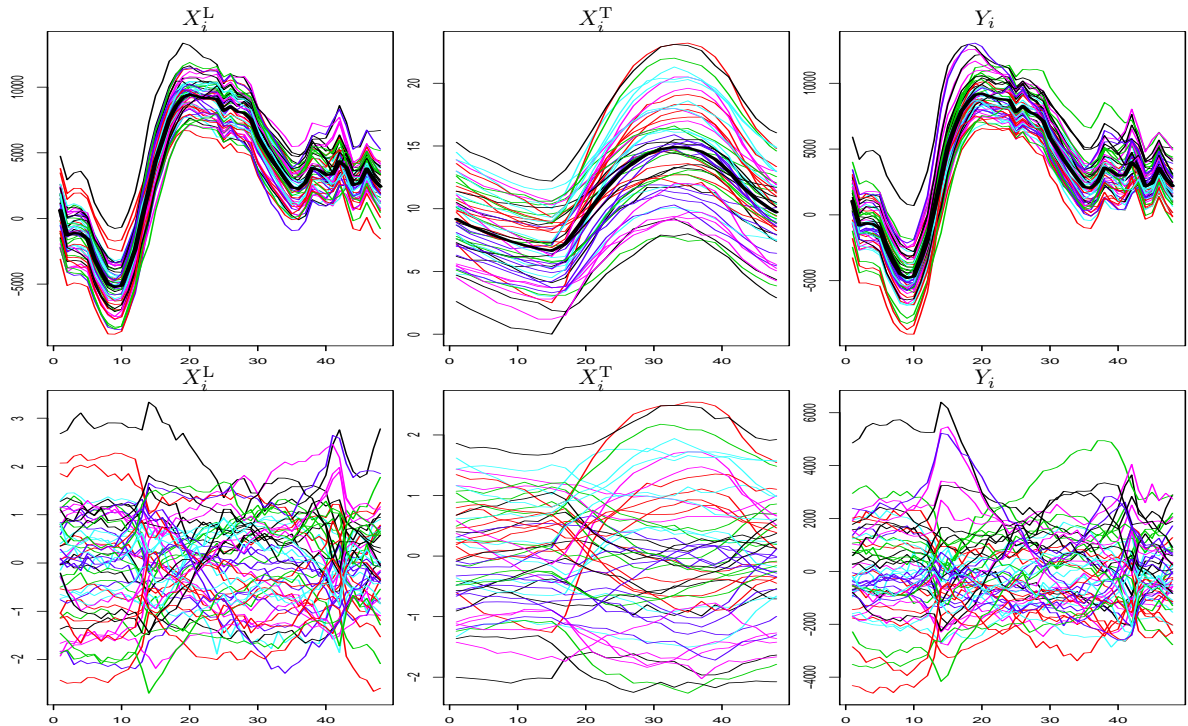, width=1\linewidth, height=3.75in} 
\end{center}
\caption{The 53 curves $X^{\mbox{\scriptsize L}}_i(\cdot)$ (top-left), $X^{\mbox{\scriptsize T}}_i(\cdot)$ (top-middle)
and $Y_i(\cdot)$ (top-bottom), together with their respective mean curves plotted together in bold black. The de-meaned and standardised curves are plotted in the bottom panels.}
\label{fig:ex:curves}
\end{figure}

To forecast the load curve on Wednesday, 2 April 2009,
we take the joined curve of the de-trended curve on Tuesday, 1 April 2009 ($=X^{\mbox{\scriptsize L}}(\cdot)$)
and the temperature curve on 2 April 2009 ($=X^{\mbox{\scriptsize T}}(\cdot)$) as the regressor,
i.e. $X(\cdot) = (X^{\mbox{\scriptsize L}}(\cdot), X^{\mbox{\scriptsize T}}(\cdot))$.
We use all the pairs of curves on Tuesday and Wednesday in April from 1996 to 2008
as our observations to fit a curve regression model, and the total number of observations is $n=53$.
As the curves $X^{\mbox{\scriptsize L}}_i(\cdot)$ range between -10000 and 10000 while
$X^{\mbox{\scriptsize T}}_i(\cdot)$ between 0 and 20, 
we apply a simple standardisation step to arrange the regressor observations in the same scale.
Those 53 pair curves $\{X_i(\cdot), \, Y_i(\cdot)\}$ are plotted in Figure~\ref{fig:ex:curves} 
together with their de-meaned and standardised counterparts.

\begin{figure}[htbp]
\psfrag{IC(q)}{$\scriptstyle{IC_2(q)}$}
\psfrag{q}{$\scriptstyle{q}$}
\begin{center}
\epsfig{file=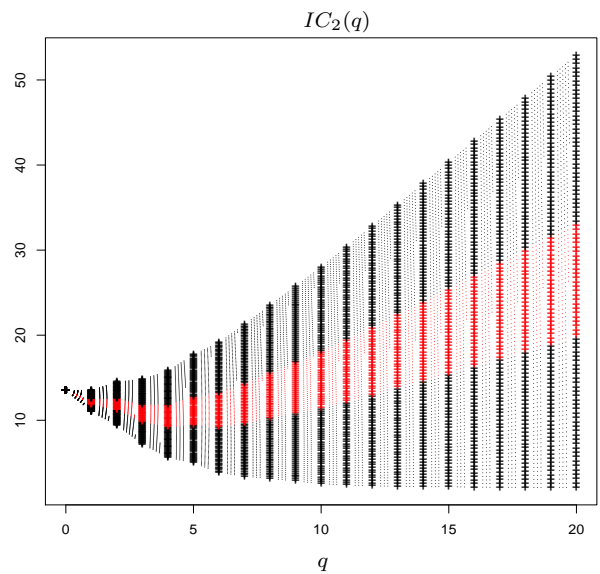, width=.5\linewidth, height=3in} 
\end{center}
\caption{Plots of $IC_2(q)$ against $q$ for 100 different values of $\tau_2$.
The curves with the minimum attained at $q=4$ are highlighted in red.}
\label{fig:ex:ic}
\end{figure}

From those observations, we form a sample covariance matrix
\begin{eqnarray} \label{q1}
\wh \Sigma (u, v) = \frac{1}{53} \sum_{i=1}^{53} \{Y_i(u) - \bar Y(u)\}\{X_i(v) - \bar X(v)\},
\end{eqnarray}
where $\bar Y(u) = \frac{1}{53} \sum_{1 \le i \le 53} Y_i(u)$ is the
average of all the de-trended daily curves on Wednesdays in April between 1996 and 2008,
and $\bar X(v)$ is obtained analogously. 
Applying the SVD to $\wh \Sigma (u, v)$, we obtain the estimators 
$(\wh \lam_k, \wh \varphi_k, \wh \psi_k)$.
To determine the correlation dimension, we apply the information criterion $IC_2(q)$ with 100 different values of $\tau_2$,
as discussed in Section~\ref{sec:estimation}.
Figure~\ref{fig:ex:ic} shows $IC_2(q)$ against $q$ for each of the 100 $\tau_2$-values. 
With this set of data, $q=4$ minimises $IC_2(q)$ over the longest interval of $\tau_2$,
which leads to the estimator $\wh r = 4$. 
Then our predicted load curve is of the form
\begin{eqnarray} \label{q2}
\wh Z(u) = \wh L_w + \bar Y(u) + \sum_{j=1}^4 \wh \xi_j \wh \varphi_j(u),
\end{eqnarray}
where $\wh L_w$ is the predicted weekly trend for the week containing 2 April 2009
from the GAM (\ref{eq:trend:two}) in Section~\ref{sec:week},
$\bar Y(u)$ is the mean curve as in (\ref{q1}),
and $\wh \xi_j, \, j=1, \ldots, 4$ are the predictors based on linear regression models defined as follows.
Based on Theorem~\ref{thm:one}, the curve linear regression $Y_i(\cdot)$ on $X_i(\cdot)$
may be recast into $\wh r=4$ ordinary regression models
\begin{eqnarray} \label{q4}
\wh \xi_{ij} = \sum_{k=1}^{10} \beta_{jk}\wh \eta_{ik} + \vep_{ij}, \quad
i=1, \cdots, 53, \;\; j=1, \cdots, 4,
\end{eqnarray}
where
\begin{eqnarray*}
\wh \xi_{ij} = \int_{\cI_1} \{Y_i(u)  - \bar Y(u)\}\wh\varphi_j(u) du, \qquad
\wh \eta_{ik} = \int_{\cI_2} \{X_i(v) - \bar X(v)\}\wh\psi_k(v) dv,
\end{eqnarray*}
see (\ref{q3}). In (\ref{q4}) we choose to use the first 10 singular value components of the regressor only,
as having more terms does not improve the forecasting result dramatically.
Based on the least squares estimators $\wh \beta_{jk}$ from the regression models (\ref{q4}), 
we obtain the predictors $\wh \xi_j$ as $\wh \xi_j = \sum_{k=1}^{10} \wh \beta_{jk} \wh \eta_k$,
where $\wh \eta_k = \int_{\cI_2} \{X(v) - \bar X(v)\} \wh \psi_k(v) dv$.

\begin{figure}[htbp]
\psfrag{2 April 2009}{\footnotesize{2 April 2009}}
\begin{center}
\epsfig{file=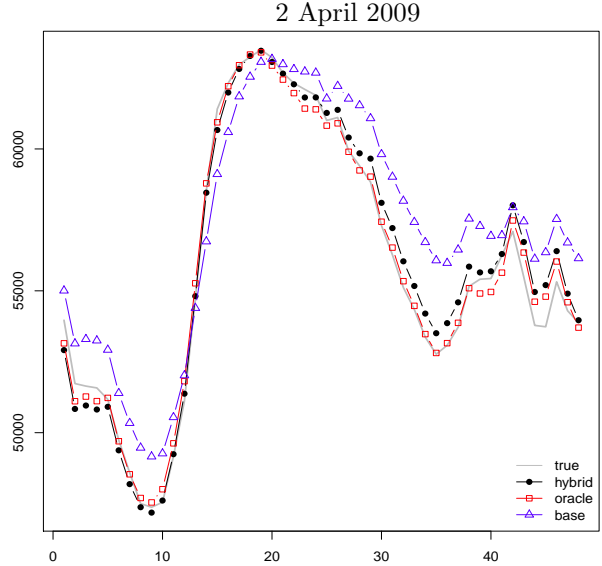, width=.5\linewidth, height=3in} 
\end{center}
\caption{The true daily load curve (grey, solid) of 2 April 2009, together with
its predicted curves by our method (black, filled circle),
the oracle method (red, empty square) and the base-line (blue, empty triangle).}
\label{fig:ex:forecast}
\end{figure}

We compare our method with two alternative predictors, the oracle and the baseline predictors. 
The oracle predictor is of the form
\begin{eqnarray} \label{q6}
\wt Z(u) = \wh L_w + \bar Y(u) + \sum_{j=1}^4 \wt \xi_j \wh \varphi_j(u),
\end{eqnarray}
which is defined similarly as our predictor (\ref{q2}) except with
$\wh \xi_j $ being replaced by $\wt \xi_j \equiv  \inner{Y(\cdot) - \bar Y(\cdot)}{\wh \varphi_j}$,
where $Y(\cdot) = Z(u) - \wh L_w$ denotes the de-trended load curve on 2 April 2009. 
Since $Y(\cdot)$ is unavailable in practice, $\wt Z(u)$ is termed as an ``oracle'' predictor.
The baseline predictor is defined as
\begin{eqnarray} \label{q7}
\bar{Z}(u) = \wh L_w + \bar Y(u),
\end{eqnarray}
which is the sum of the first two terms in our predictor, ignoring the dynamic dependence between days. 
We compare the performance of the three predictors in terms of the following two error measures
\begin{eqnarray*}
{\rm MAPE} = \frac{1}{48}\sum_{j=1}^{48} \left\vert \frac{\wh f_j - f_j}{f_j} \right\vert \quad 
{\rm and} \quad
{\rm RMSE} = \left\{\frac{1}{48} \sum_{j=1}^{48} (\wh f_j - f_j)^2\right\}^{1/2},
\end{eqnarray*}
where $\wh f_j$ and $f_j$ denote the predicted and the true loads in the $j$-th half-hour interval.
The MAPE and RMSE for our predictor $\wh Z(\cdot)$, the oracle predictor $\wt Z(\cdot)$ and
the baseline predictor $\bar{Z}(\cdot)$ are (0.91\%, 634MW),
(0.60\%, 420MW) and (3.14\%, 1911MW), respectively. 
The three predicted curves are plotted in Figure~\ref{fig:ex:forecast} together with the true curve. 
Our predictor $\wh Z(\cdot)$, making good use of the dynamic dependence across different days, 
is a significant improvement from the baseline predictor $\bar{Z}(\cdot)$.
While the oracle predictor $\wt Z(\cdot)$ is impractical as $\wt \xi_j$ is unavailable in practice, 
its superior performance in terms of both MAPE and RMSE indicates that 
the dimension reduction achieved via SVD retains the relevant dynamic information in the system. 

We briefly discuss the extension to multi-step ahead predictions using the hybrid approach,
which straightforwardly translates to producing multi-step ahead predictions from the GAM at the weekly level,
and from the ordinary (scalar) linear regression models at the daily level. 
Specifically, if the corresponding week of the multi-step ahead forecast is different from that of the one-step ahead forecast,
the forecast is obtained by plugging the average temperature and cloud cover of the week into the fitted GAM.
At the daily level modelling, the forecast of the next day's load replaces (part of) the regressor curve 
to produce that of the following day, and this is repeated until the desired multi-step ahead prediction is achieved.
In the above example, when making a two-day ahead prediction for Thursday, 3 April 2009 on 1 April 2009, 
the first part of the regressor curve becomes the predicted load curve on 2 April 2009,
while the second part is the daily temperature curve on 3 April 2009.
The two-step ahead forecast obtained following the identical steps described in this section
achieves MAPE 1.06\% and RMSE 657MW.
In general, the performance of multi-step ahead forecasts is worse than that of one-day ahead forecasts
as the errors in the latter are carried over to the errors in the former.

\section{Predicting daily loads in 2009}
\label{sec:predict2009}

To compare different predictive models more systematically, and 
to gain further appreciation of the performance of our method over different periods of a year, 
we predict the daily load curves for all days in 2009.
For each day in 2009, we use the data from 1 January 1996 to its previous day to build the prediction models
in the same manner as described in Section~\ref{sec:illustration}, i.e.  
first the trend component (i.e. as $\wh L_w$ in (\ref{q2})) is predicted by the GAM model in (\ref{eq:trend:two}), 
and then the residual process is divided into daily curves for curve linear regression. 

\subsection{Classification of daily curves}
\label{sec:class}

\begin{figure}[htbp]
\centering
\epsfig{file=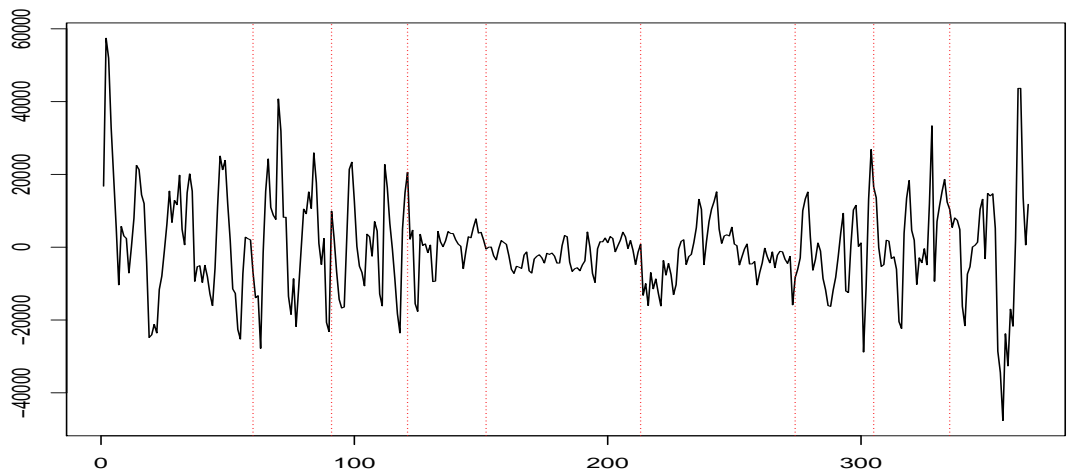, width=1\linewidth, height=3in} 
\caption{Decomposition of the daily curves from 2008 with respect to the first principal component
estimated from the pooled daily curves between 1996 and 2008:
seasonal segments are denoted by dotted, red lines.}
\label{fig:five}
\end{figure}

Discussions in Section \ref{sec:illustration} indicate that 
we need to treat the daily residual curves on each day of a week differently.
For the French electricity load dataset, we are furnished with the
\emph{day type} of each day, which is a classification of the daily curves determined by the experts at EDF.
The day type is defined with respect to different days of a week, 
and bank holidays are assigned to separate day types according to their profiles.
See Table \ref{table:daytype} for the summary of day types.
\begin{table}
\caption{Day types furnished by the EDF experts.}
\label{table:daytype}
\centering
\small{
\begin{tabular}{c|c|c|c|c|c|c|c|c}
\hline
\hline
index & 0 & 1 & 2 & 3 & 4 & 5 & 6 & 7 \\
\hline
day type & Mon & Tue--Thu & Fri & Sat &
Sun (rest) & Sun (Jun-Jul) & Sun (Aug) & Sun (Dec)
\\
\hline
\end{tabular}}
\end{table}
Furthermore, to take into account the seasonal changes 
which may be present in the shapes ($\E\{Y_i(\cdot)\}$ and $\E\{X_i(\cdot)\}$) as well as 
the dependence structure ($\Sigma(u, v)=\cov(Y_i(u), X_i(v))$) of daily curves,
we divide one year into 9 seasonal segments:
January to February, March, April, May, June to July, August to September, October, November, and December.
This segmentation was determined by inspecting the decomposition of electricity loads
with respect to adaptively chosen orthonormal functions.
More precisely, we performed principal component analysis on the pool of de-meaned daily curves (according to the day type),
and decomposed them with respect to the first principal direction.
By examining the changes in the decomposition over a year (see Figure \ref{fig:five})
we obtained the segmentation of a year as provided above. 

While the above classification lacks a rigorous statistical ground,
the prediction model based on this classification performs well in practice.
Besides, classification of electricity load curves can stand alone as an independent research problem
which has attracted considerable attention, see e.g. 
\citet{chiou2007}, \citet{ray2006}, \citet{serban2005} and \citet{james2003} for functional clustering,
and \citet{Anto10} in the context of electricity loads classification.
In summary, each daily curve is classified according to the day of a week and the season of a year,
and there are 67 pairs of classes for any two consecutive days between 1996 and 2009.
For each pair of classes, we fit a prediction model separately in the same manner as described in Section~\ref{sec:illustration}.

\subsection{Prediction comparisons}
\label{sec:sim}

In applying the proposed hybrid method, we consider four different versions H1--H4 depending on the choice of regressor.
H1 uses the load curve on the current day as the regressor (i.e. $X(\cdot) = X^{\mbox{\scriptsize L}}(\cdot)$).
H2 uses the joined curve of the load curve on the current day and the temperature curve on the next day 
(i.e. $X(\cdot) = (X^{\mbox{\scriptsize L}}(\cdot), X^{\mbox{\scriptsize T}}(\cdot))$), 
as it has been practiced in Section \ref{sec:illustration}.
H3 adopts the same regressor as H1 but with a half-day curve such that,
if we are forecasting the electricity load from 00:30 to 12:00 on the next day, 
the load curve on the current day from 12:30 to 24:00 is used as the regressor curve;
when forecasting the curve from 12:30 to 24:00, the regressor curve is the load curve
from 00:30 to 12:00 on the same day.
Similarly, H4 employs the same regressor as H2 but also with half-day curves. 
To facilitate a more comprehensive comparison, we predict the daily load curves
by our proposed hybrid method (\ref{q2}), the oracle method (\ref{q6}) 
and the baseline method (\ref{q7}). 
We also include in the comparison study, the prediction results from the EDF operational model, 
the seasonal ARIMA model (denoted as SARIMA) as in \citet{Tayl08},
a combination of GAM and SARIMA (GSARIMA) method, 
and the exponential smoothing technique (EST) discussed in \citet{taylor2010}. 
In total, there are 10 different models used in our comparison study.

\begin{figure}[htbp]
\begin{center}
\begin{tabular}{c}
\epsfig{file=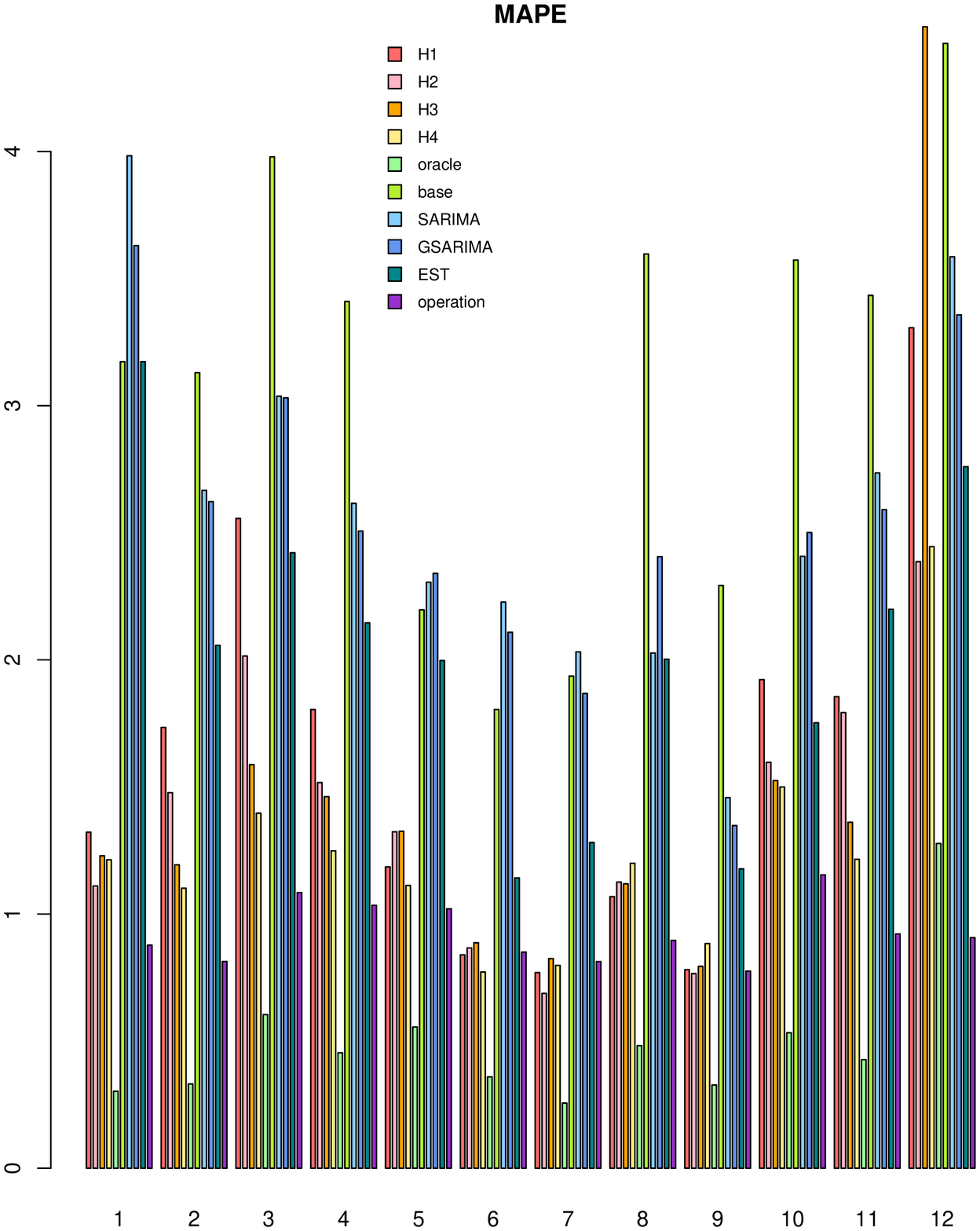, width=1\linewidth, height=3.55in} \\
\epsfig{file=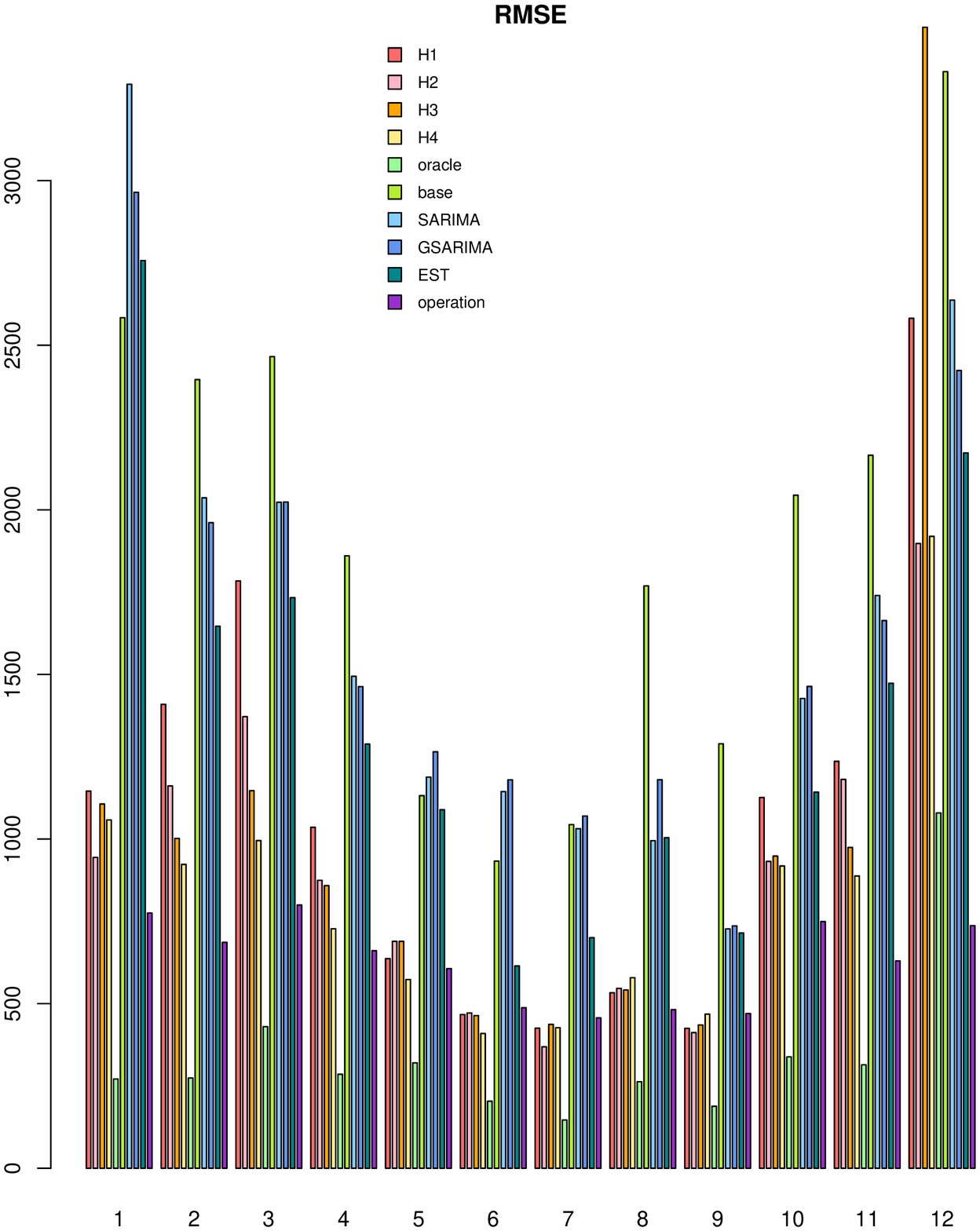, width=1\linewidth, height=3.55in} 
\end{tabular}
\end{center}
\caption{\footnotesize{Bar plots of MAPE (top) and RMSE (bottom) with respect to months.}}
\label{fig:sim:ss}
\end{figure}

\begin{figure}[htbp]
\begin{center}
\begin{tabular}{c}
\epsfig{file=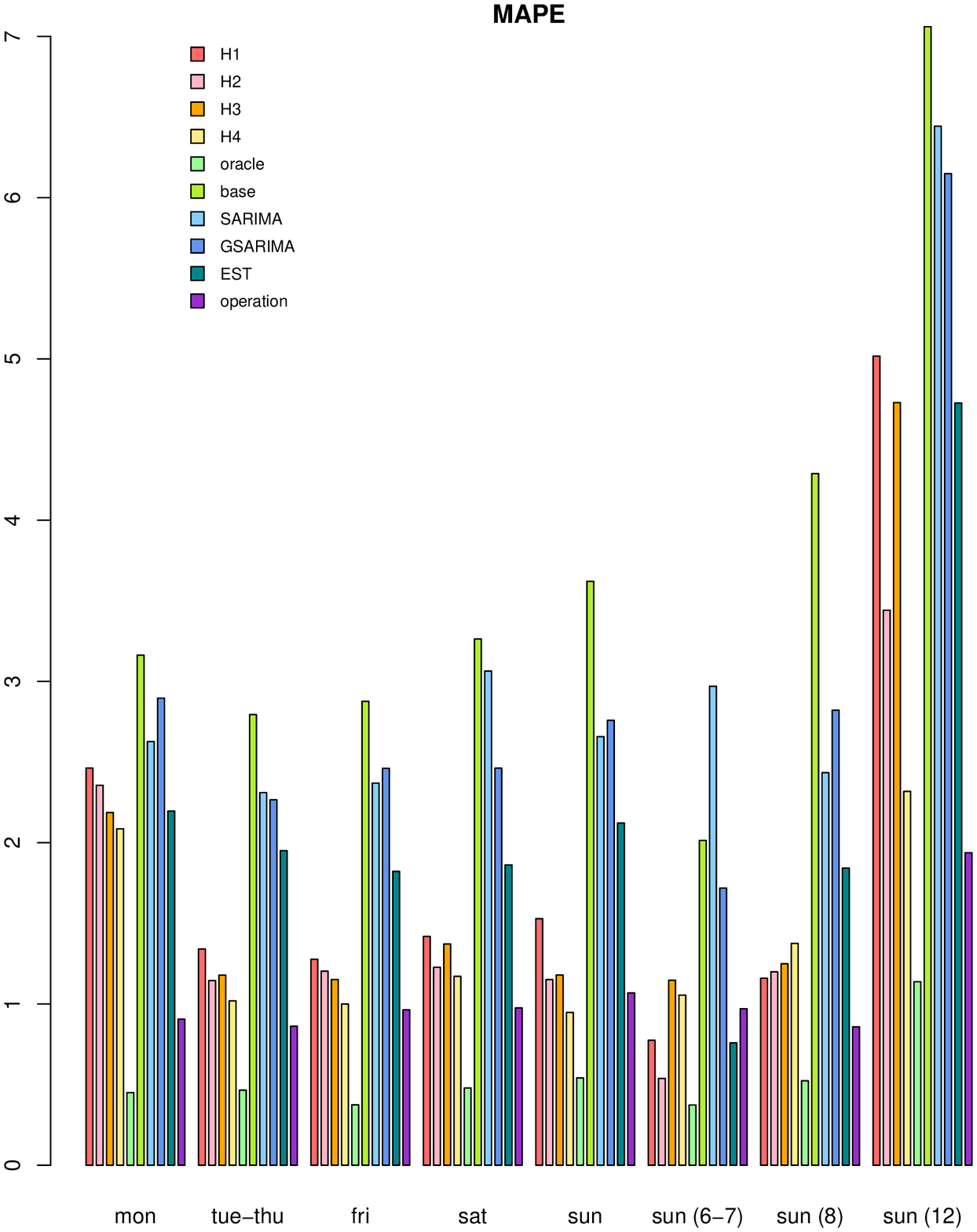, width=1\linewidth, height=3.55in} \\
\epsfig{file=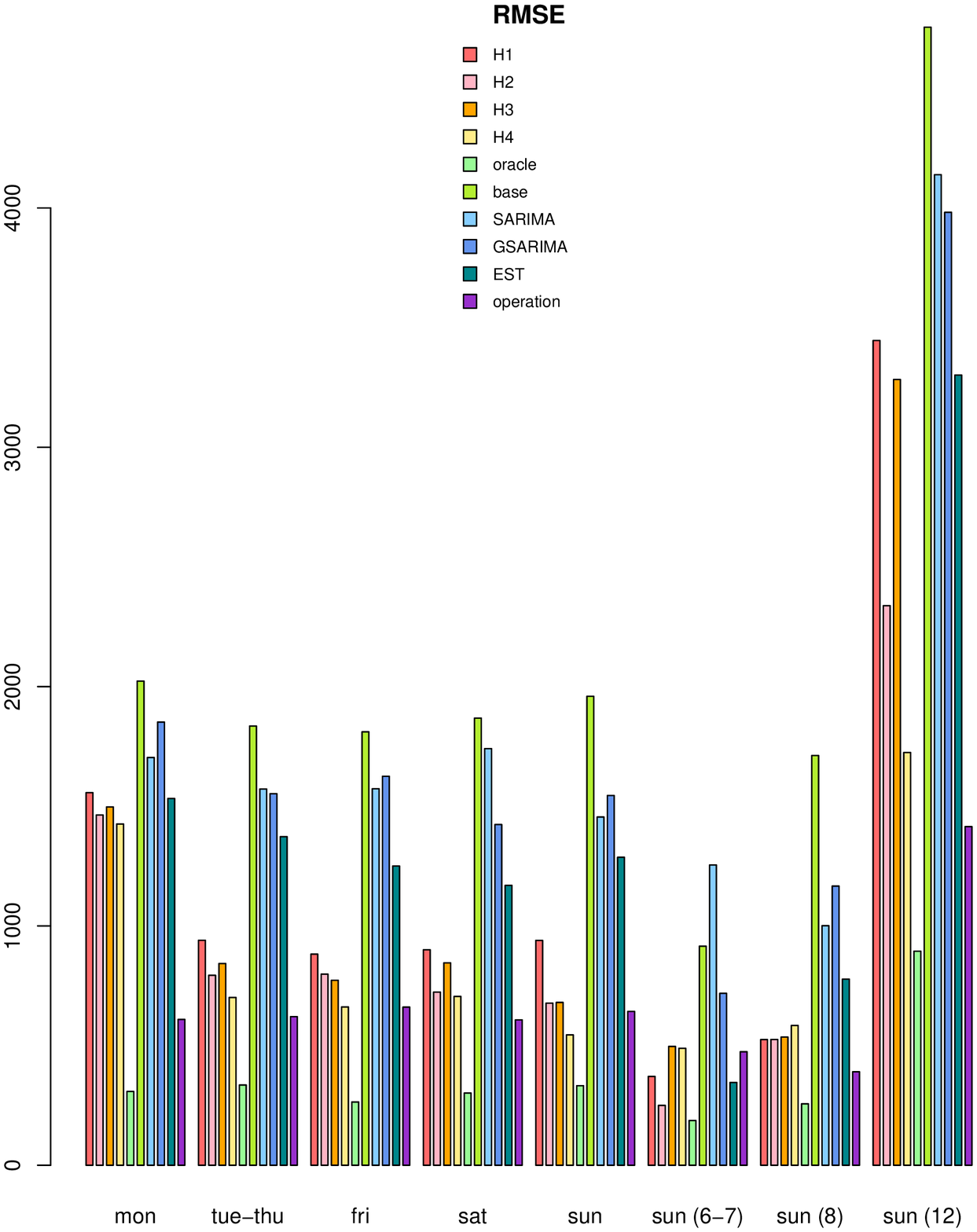, width=1\linewidth, height=3.55in} 
\end{tabular}
\end{center}
\caption{\footnotesize{Bar plots of MAPE (top) and RMSE (bottom) with respect to the day type determined by experts;
from left to right: Mondays, Tuesdays--Thursdays, Fridays, Saturdays, Sundays (except for June--August and December), Sundays in June--July, Sundays in August and Sundays in December.}}
\label{fig:sim:dt}
\end{figure}

Denote the number of observations for each pair of classes by $n$.
Since we impose an upper bound of 10 on the correlation dimension $r$, 
we choose to include those classes with $n$ greater than 15 in our comparison study.
Also, only the first 10 $\eta_{ik}$s are used in the scalar linear regression models (\ref{multiple:lm}),
as having more than 10 terms does not improve the results dramatically while $n$ is allowed to be as small as 15.
We further note that it is considered a more challenging task 
to forecast electricity loads for holidays than those for working days,
and often additional prior information is used for holidays in practice.
Instead of making the whole exposition over-complicated, we focus on the forecasting for the working days only.
There are 315 days in total where all the conditions stated above are satisfied.
Note that in the hybrid approach, we require the forecasts of the 
average temperature of the following week, as well as the temperature curve of the next day.
As such information can easily be furnished by M\'{e}t\'{e}o-France for this particular dataset,
we may assume that the forecast of the next day's temperature has been provided in the form of a curve,
and the weekly average temperature of the following week can be replaced by the mean of such a forecast
(in accordance with the assumption that the long-term trend to vary little within each week).
Since the resulting MAPE (1.38\%) and RMSE (891MW) from (H2) with the predicted temperature values
are only slightly worse than those obtained with the true temperature values (MAPE 1.35\%, RMSE 869MW),
we report in what follows the results obtained assuming that all such necessary information is available.
Forecasting errors measured by the MAPE and RMSE are summarised in Table \ref{table:one}, and
we also present the errors with respect to different seasons and day types
in Figures \ref{fig:sim:ss}--\ref{fig:sim:dt}.
\begin{table}
\caption{Summary of MAPE and RMSE of the electricity load forecasts for 01/01/2009--31/12/2009
from our hybrid modelling (H1, H2, H3, H4), oracle, base, SARIMA, GSARIMA, EST and operational model.}
\label{table:one}
\centering
\footnotesize{
\begin{tabular}{c|c|c|c|c|c|c|c|c|c|c}
\hline
\hline
& H1 & H2 & H3 & H4 & oracle & base & SARIMA & GSARIMA & EST &operation \\
\hline
MAPE ($\%$) & 1.54 & 1.35 & 1.37 & 1.20 & 0.46 & 3.05 & 2.55 & 2.49 & 1.97 & 0.93 \\
RMSE (MW) & 1018 & 869 & 918 & 787 & 317 & 1882 & 1607 & 1586 & 1330 & 625 \\
\hline
\end{tabular}}
\end{table}

The prediction based on any model considered is more accurate in summer than in winter, see Figure \ref{fig:sim:ss}.
The relative difficulty of load forecasting in winter has been noted for the French dataset
in \citet{Dord08}, \citet{Dord11} and \citet{cugliari2011}.
SARIMA and GSARIMA are consistently outperformed by other methods by a large margin,
and between the two, GSARIMA achieves smaller forecasting errors.
Between H1 (H3) and H2 (H4), the latter attains considerably smaller forecasting errors
as it makes use of more information on the temperature,
although Figure \ref{fig:sim:ss} shows that this observation is not held consistently throughout the year.
We note that the performance of our approach may further be improved by 
making an adaptive choice of regressor curve dependent on the level of temperature.

From Figure \ref{fig:sim:ss}, it is interesting to observe that the half-day based approaches, H3 or H4, 
achieve better forecasting performance than H1 or H2 in some colder months (February--April, October--November), 
while the opposite is true in warmer months.
This may be understood in relation with the variability among the curves, 
which is considerably greater in winter than in summer (see e.g. Figure \ref{fig:four}). 
On a similar note, while forecasting errors from the EDF operational model are smaller than 
those from hybrid approaches on average, the difference is noticeably reduced from May to September.
Indeed, H1 and H2 return errors which are comparable to or even smaller than those from the operational model 
in June, July and September.
In terms of day type, the forecasting errors from the hybrid methods 
are larger on Mondays than for the rest of a week on average (see Figure \ref{fig:sim:dt}),
which may also be due to the greater variability in the relationship between the curves from Sundays and Mondays.
The oracle predictor attains the minimum errors throughout the year except for in December, which suggests that 
there is a scope for improvement in the hybrid approach by improving
the linear regression fit at the daily level.

There are certain factors which are known to have substantial influence on daily electricity loads 
yet have not been incorporated into our hybrid modelling.
For example, from November to March, EDF offers special tariff days to large businesses as financial incentives,
which are activated to cut heavy electricity consumption in winter.
Since the scheme is known to affect not only the daily loads on the special tariff days but also 
on the days before and after those days, we expect that including prior information on such days, 
e.g. by creating new classes, can further improve the quality of the forecasts especially in winter.

\section{Conclusions}
\label{sec:conclusion}

In this paper, we proposed a hybrid approach to electricity load modelling
with the aim of forecasting daily electricity loads.
In the hybrid procedure, we model the overall and seasonal trends of the electricity load data at the weekly level,
by fitting a GAM with temporal and meteorological factors as explanatory variables.
At the daily level, the serial dependence among the daily load curves is modelled
under the assumption that the curves from two successive days have a linear relationship,
and we propose a framework which effectively reduces the curve linear regression 
to a finite number of scalar linear regression problems. 
To the best of our knowledge, it has not been explored elsewhere 
to model the multi-layered features of electricity load dataset at multiple levels separately.
Compared to the current operational model at EDF, our proposed method is more model-centred and 
developed without much of the specific knowledge that have been included in the former,
while it still retains a competitive prediction capacity.
We also note that our approach has the potential to be more adaptive to changing electricity consumption environment,
as well as being applicable to a wider range of problems without much human intervention.

When applying the hybrid approach to real-life dataset in Section \ref{sec:sim}, 
some factors which may have substantial influence over daily electricity loads have not been fully exploited.
This could have resulted in worsening the performance of our method for winter days when compared to the operational model,
and it remains as a task to incorporate such relevant information into our method for practical applications. 
Also, as briefly mentioned in Section \ref{sec:sim}, an adaptive choice of the regressor curve, 
depending e.g. on the level of temperature, may lead to better results in daily load forecasting.
Indeed, an automatic selection of the regressor in the curve linear regression framework may benefit 
the prediction performance as a generic tool beyond the electricity load forecasting, 
and we leave the problem for future research.

\bibliographystyle{apalike}
\bibliography{fbib}
\end{document}